\documentclass[conference]{IEEEtran}
\usepackage{cite}
\usepackage{amsmath,amssymb,amsfonts}
\usepackage{algorithmic}
\usepackage{graphicx}
\usepackage{tabularx}
\usepackage{textcomp}
\usepackage{xcolor}
\usepackage{stfloats}
\usepackage{multirow}
\def\BibTeX{{\rm B\kern-.05em{\sc i\kern-.025em b}\kern-.08em
    T\kern-.1667em\lower.7ex\hbox{E}\kern-.125emX}}
\begin{document}

\title{A 3D Non-Stationary Geometry-Based Stochastic Model for Industrial Automation Wireless Communication Systems}

\author{Yuxiao Li\textsuperscript{1,2}, Cheng-Xiang Wang\textsuperscript{1,2*}, Yang Liu\textsuperscript{1,3}
\\
\textsuperscript{1}National Mobile Communications Research Laboratory, School of Information of Science and Engineering,
\\Southeast University, Nanjing 210096, China.\\

\textsuperscript{2}Purple Mountain Laboratories, Nanjing 211111, China.\\
\textsuperscript{3}JiangNan University, Wuxi 214122, China.\\
\textsuperscript{*}Corresponding Author: Cheng-Xiang Wang
\\
Email: yuxli@seu.edu.cn, chxwang@seu.edu.cn, ly71354@163.com
}
\maketitle
\begin{abstract}

Industrial automation is one of the key application scenarios of the fifth generation (5G) wireless communication network. The high requirements of industrial communication systems for latency and reliability lead to the need for industrial channel models to support massive multiple-input multiple-output (MIMO) and millimeter wave communication. In addition, due to the complex environment, huge communication equipment, and numerous metal scatterers, industrial channels have special rich dense multipath components (DMCs). Considering these characteristics, a novel three dimensional (3D) non-stationary geometry-based stochastic model (GBSM) for industrial automation wireless channel is proposed in this paper. Channel characteristics including the transfer function, time-varying space-time-frequency correlation function (STFCF), and root mean square (RMS) delay spread, model parameters including delay scaling factor and power decay factor are studied and analyzed. Besides, according to the indoor factory scenario classification of the 3rd
Generation Partnership Project (3GPP) TR 38.901, two sub-scenarios considering the clutter density are simulated. Simulated cumulative distribution functions (CDFs) of RMS delay spread show a good consistency with the measurement data.
\end{abstract}
\section{Introduction}
Many new technologies have been proposed in 5G network, bringing new characteristics to the wireless channel. A survey of 5G channel measurements and modeling was given in \cite{[31]}, describing the new requirements for 5G channel modeling. Besides, as a key application of 5G and the sixth generation (6G) network, channel modeling for industrial automation is very important\cite{[6G],[1]}.

In the future 6G wireless communication system, it is urgent for the industry to build the Industrial Internet of Things (IIoT) wireless network to connect the equipment, people, and data, improving data transmission and facilitating industrial revolution.
It has been reported that massive MIMO and beam-forming technologies are expected to apply to the industrial scenario \cite{[3GPP]}. Furthermore, more spectrum resources may be available in IIoT as the center frequency expands into the millimeter wave band.
Besides, compared with other scenes, the channel in industry has new channel characteristics such as changing path loss, rich scatterers, and multi-mobility due to the existence of a large number of mechanical equipments, metals, and sensors \cite{[31],[1]}.
So much work had found that the channel impulse response in industrial scenario can be modeled as a summary of specular multipath components (SMCs) and DMCs\cite{[IF1],[IF2]}.
DMCs account for a larger proportion in an indoor factory \cite{[IF1],[IF2]}. The number of multipath components required to capture a majority of the energy is quite large.
Experimental analysis of DMCs in an industrial environment was also made based on measurements in \cite{[dmc_mea]}.
To sum up, an elaborate channel model for IIoT needs to support not only massive MIMO, millimeter wave bands, but also the special channel characteristics of IIoT.

The SMCs result from the specular reflections from large physical objects. Meanwhile, those objects with smaller size compared to the wavelengths or rough surfaces will lead to scattering, producing a large number of weak paths, which are identified as the DMCs. In the beginning, room electro-magnetics (RM) theory \cite{[RM]} studied various aspects of diffuse wideband microwave propagation in a room in analogy with the established discipline of room acoustics (RA) theory.
Later, in order to describe the channel DMCs, reference \cite{[DMC1]} attached a dense multi-path cluster to each coherent component, reference \cite{[IF1]} modeled the power of DMCs as a function of delay and reservation time, and the author of \cite{[DMC3]} also proposed a DMC add-on model for COST2100 channel model. Besides, the propagation characteristics of DMCs were analyzed in \cite{[DMC4]} and a method to add DMCs on GBSM was also proposed.

Some channel models have been proposed for industrial channel so far, including ray tracing channel models, statistical models, and GBSMs\cite{[26],[liuyang]}. Besides, a survey of 3GPP TR 38.901 (call 3GPP for short below) standardized 5G channel model \cite{[26]} for IIoT was given in \cite{[3GPP]}, giving four IIoT sub-scenarios and the models of channel parameters, including the path loss and the line-of-sight (LOS) probability, the RMS delay spread, and the angular spread. However, these models cannot support either massive MIMO and millimeter wave communication or the DMCs in IIoT. A general 5G channel model was proposed in \cite{[29]}, but it also did not consider the DMCs modeling.

In order to describe the rich DMCs in the industrial scenario and meet the requirements of IIoT channel model, this paper has proposed a novel comprehensive 3D non-stationary GBSM for industrial automation wireless communication systems. The model supports massive MIMO, millimeter wave bands and has the ability to describe the rich DMCs in IIoT.
The rest of this paper is organized as follows. Section~\uppercase\expandafter{\romannumeral2} introduces the proposed GBSM for industrial scenario. Section~\uppercase\expandafter{\romannumeral3} lists some important characteristics of the model. Next, the simulation results and analysis of the model are presented in Section~\uppercase\expandafter{\romannumeral4}. Finally, conclusions are drawn in Section~\uppercase\expandafter{\romannumeral5}.
\section{A non-stationary Channel Model for Industrial scenario}

\begin{figure}[tb]
	\centerline{\includegraphics[width=0.5\textwidth]{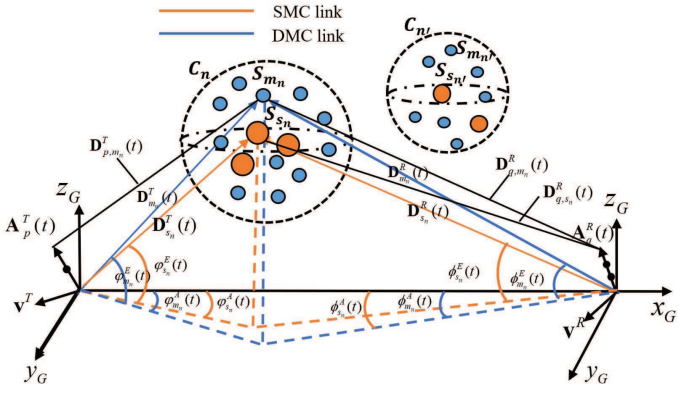}}
	\caption{A novel 3D non-stationary GBSM for IIoT.}
	\label{fig_1}
\end{figure}
Let us consider an $M_T\times M_R$ MIMO communication system, where $M_T$ and $M_R$ are the number of antennas of transmitter (Tx) and receiver (Rx), respectively. Let $A_q^R$ represent the receiving antenna $q$, $A_p^T$ represent the transmitting antenna $p$. Considering real industries, there might be many movable scatterers around them. Tx and Rx can be also moving. It has been observed that the angular (delay) power spectrums of the DMCs are not angular-white (delay-white), but significantly correlated with the properties of the specular part of the channel. In the proposed model, those scatterers producing DMCs are assumed to have smaller size and be distributed near those producing SMCs. They are distributed in clusters. In each cluster $C_n$, the scatterer producing $s_n$-th SMC is denoted as $S_{s_n}$, while that producing $m_n$-th DMC is denoted as $S_{m_n}$, as shown in Fig.~1.

The scattering environments are modeled as effective clusters\cite{[29]}, thus the total link delay of $C_n$ consists of the delay from Tx to $C_n$ (the first bounce), the delay from Rx to $C_n$ (the last bounce), and the virtual link delay between the first bounce and the last bounce. Besides, clusters might be unobservable to Tx or Rx, which makes Rx cannot receive those signal components. For simplify, all the scatterers in a cluster are assumed to have the same moving speed and visibility. The geometry parameters are listed in Table~\ref{tab1}.
\begin{table*}[tb]
 	\centering
 	\caption{Definition of Key Parameters.}
 		\begin{tabular}{|c|c|}			
			\hline
 			\textbf{Parameters}&\textbf{Definition}\\
 			\hline
            $\psi_A^R$, $\psi_E^R$&Azimuth and elevation angles of the receive array
  			\\\hline
  			$\psi_A^T$, $\psi_E^T$&Azimuth and elevation angles of the transmit array
  			 \\\hline
  			 $\phi_{s_n}^A(t)$, $\phi_{s_n}^E(t)$&Azimuth and elevation angles between $S_{s_n}$ and the receive array center
  			  \\ \hline
  			 $\varphi_{s_n}^A(t)$, $\varphi_{s_n}^E(t)$&Azimuth and elevation angles between $S_{s_n}$ and the transmit array center
  			  \\ \hline
  			 $\phi_{m_n}^A(t)$, $\phi_{m_n}^E(t)$&Azimuth and elevation angles between $S_{m_n}$ and the receive array center
  			  \\ \hline
  			 $\varphi_{m_n}^A(t)$, $\varphi_{m_n}^E(t)$&Azimuth and elevation angles between $S_{m_n}$ and the transmit array center
 			  \\\hline
  			 ${\bf{A}}_p^T(t)$, ${\bf{A}}_q^R(t)$&3D position vectors of $A_p^T$ and $A_q^R$
   			\\ \hline
  			  $\bf{D}$&3D position vector of receive array center
   			\\  \hline
  			${\bf{D}}_{pq}^{LoS}(t)$ & 3D distance vector of the LOS component between $A_p^T$ and $A_q^R$
 			\\\hline
 			${\bf{D}}_{s_n}^R(t)$, ${\bf{D}}_{s_n}^T(t)$ & 3D distance vector between $S_{s_n}$ and the receive (transmit) array center
            \\\hline
 			${\bf{D}}_{q,s_n}^R(t)$, ${\bf{D}}_{p,s_n}^T(t)$ & 3D distance vector between $S_{s_n}$ and $A_p^T$ ($A_q^R$)
  			\\\hline
 			${\bf{D}}_{m_n}^R(t)$, ${\bf{D}}_{m_n}^T(t)$ & 3D distance vector between $S_{m_n}$ and the receive (transmit) array center
   			\\\hline
 			${\bf{D}}_{q,m_n}^R(t)$, ${\bf{D}}_{p,m_n}^T(t)$ & 3D distance vector between $S_{m_n}$ and $A_p^T$ ($A_q^R$)
 			\\\hline
 		    $\textbf{v}^R$, $\textbf{v}^T$&3D velocity vector of receive (transmit) array
 			\\\hline
            $\textbf{v}_n^R$, $\textbf{v}_n^T$&3D velocity vector of the first bounce and the last bounce of $C_n$ (including $S_{s_n}$ and $S_{m_n}$)
	\\\hline
 		\end{tabular}
 		\label{tab1}
 \end{table*}
%

\subsection{Channel Impulse Response}
After determining the geometric framework, the model can be characterized by an $M_R\times M_T$ matrix $\textbf{H}(t,\tau)=[h_{qp}(t,\tau)]_{M_R\times M_T}$, where $h_{qp}(t,\tau)$ is the channel impulse response between $A_q^R$ and $A_p^T$ and consists of LOS and non-LOS (NLOS) components. It is computed as
\begin{flalign}\label{eq_CIR}
     \begin{split}
h&_{qp}(t,\tau)=\sqrt{\frac{K}{K+1}}h_{qp}^{LOS}(t)\delta[(\tau-\tau^{LOS}(t))]\\
&+\sqrt{\frac{1}{K+1}}\sum_{n=1}^{N(t)}\sum_{s=1}^{{S_n}(t)}h_{qp,{s_n}}^{SMC}(t)\delta[\tau-\tau_{s_n}^{SMC}(t)]\\
&+\sqrt{\frac{1}{K+1}}\sum_{n=1}^{N(t)}\sum_{m=1}^{M_n(t)}h_{qp,m_n}^{DMC}(t)\delta[\tau-\tau_{m_n}^{DMC}(t)]
     \end{split}
\end{flalign}
where the first item is the LOS component, the second and third items are SMCs and DMCs, respectively. The symbol $K$ is the Rician factor, $N(t)$ represents the number of clusters, $S_n(t)$ represents the random number of scatterers producing SMCs, and $M_n(t)$ represents the random number of smaller scatterers producing DMCs. They are assumed to follow Poisson distribution with an arrival rate of $\tilde{\lambda}_{smc}$ and $\tilde{\lambda}_{dmc}$. According to the experimental analysis of measurement results in an industry in \cite{[dmc_mea]} and the standardized model parameters \cite{[26]}, the mean of $S_n(t)$ and $M_n(t)$ are set 3 and 17 respectively. They are given in Table~\ref{tab2}.

The channel coefficients of LOS component $h_{qp}^{LOS}(t)$, and DMC $h_{qp,m_n}^{DMC}(t)$ are computed as
\begin{equation}
\label{h_LOS}
\begin{split}
&h_{qp}^{{LOS}}(t)=\\
&\left[
\begin{matrix}
F_{p,V}^T({\bf{D}}_{qp}^{LOS}(t),\bf{A}_p^T(t))\\
F_{p,H}^T({\bf{D}}_{qp}^{LOS}(t),\bf{A}_p^T(t))
\end{matrix}\right]^T
\left[
\begin{matrix}
e^{j\Phi_{\text{LOS}}}&0\\
0&-e^{j\Phi_{\text{LOS}}}
\end{matrix}
\right]\\
&\left[
\begin{matrix}
F_{q,V}^R({\bf{D}}_{qp}^{LOS}(t),\bf{A}_q^R(t))\\
F_{q,H}^R({\bf{D}}_{qp}^{LOS}(t),\bf{A}_q^R(t))
\end{matrix}\right]
\sqrt{P_{LOS}(t)/K}e^{-j2\pi f_{qp}^{\text{LOS}}(t)t}
\end{split}
\end{equation}
\begin{equation}
\begin{split}
&h_{qp,m_n}^{DMC}(t)=\\
&\left[
\begin{matrix}
F_{p,V}^T({\bf{D}}_{m_n}^{T}(t),\bf{A}_p^T(t))\\
F_{p,H}^T({\bf{D}}_{m_n}^{T}(t),\bf{A}_p^T(t))
\end{matrix}\right]^T
\left[
\begin{matrix}
e^{j\Phi_{m_n}^{VV}}&\sqrt{\kappa_{m_n}}e^{j\Phi_{m_n}^{VH}}\\
\sqrt{\kappa_{m_n}}e^{j\Phi_{m_n}^{HV}}&e^{j\Phi_{m_n}^{HH}}
\end{matrix}
\right]\\
&\left[
\begin{matrix}
F_{q,V}^R({\bf{D}}_{m_n}^{R}(t),\bf{A}_q^R(t))\\
F_{q,H}^R({\bf{D}}_{m_n}^{R}(t),\bf{A}_q^R(t))
\end{matrix}\right]
\\&
\sqrt{P_{m_n}^{DMC}(t)}e^{-j2\pi (f_{q,m_n}^{R}(t)+f_{p,m_n}^{T}(t))t}
\end{split}
\end{equation}

The channel coefficient of SMC $h_{qp,{s_n}}^{SMC}(t)$ is similar to DMC and can be obtained by replacing $m_n$ with $s_n$. The superscripts V and H denote vertical polarization and horizontal polarization respectively, functions $F^T(\textbf{a},\textbf{b})$ and $F^R(\textbf{a},\textbf{b})$ are antenna patterns with input vectors $\textbf{a}$ and $\textbf{b}$, the phases of LOS and NLOS $\Phi_c^d$ are uniformly distributed within $(0,2\pi)$, and $\kappa$ is the cross polarization ratio. 
The Doppler frequency shift due to the movement of the receiver and $S_{s_n}$ is calculated as $f_{q,{s_n}}^R(t) = {\langle {\bf{D}}_{q,{s_n}}^R(t),{\bf{v}}^R-{\bf{v}}_n^R \rangle}/[{\lambda_c \left\| {\bf{D}}_{q,{s_n}}^R(t) \right\|}]$, where $\langle \cdot,\cdot \rangle$  is the inner product, $\left\| \cdot \right\|$ calculates the Frobenius norm.
Similarly, other frequency shifts $f_{q,{s_n}}^T(t)$, $f_{qp}^{\text{LOS}}(t)$, $f_{q,m_n}^{R}(t)$ and $f_{p,m_n}^{T}(t)$ can be also obtained easily. The power of LOS component $P_{LOS}(t)$, SMC $P_{s_n}^{SMC}(t)$, and DMC $P_{m_n}^{DMC}(t)$ will be introduced in the following sub-section.

\subsection{SMCs and DMCs Modeling}
\subsubsection{Angular Domain}

 Assume that the angles between the cluster center and the transmitter or receiver follow wrapped Gaussian distributions\cite{[29]}, whose standard deviations can be denoted by [std($\phi_n^A$), std($\phi_n^E$), std($\varphi_n^A$), std($\varphi_n^E$)] and are determined by parameter estimation. The travel angels of $S_{s_n}$ and $S_{m_n}$ are obtained by adding a Laplace distribution random angular offset with zero mean and standard deviation of 1 degree (0.017 radian) at the basis of clusters' travel angles. Besides, the study on the propagation characteristics of DMCs \cite{[dmc_mea]} found that DMCs have larger angular spreads than SMCs. The standard deviation of angular offset of DMC is a scenario-dependent constant, whose typical value includes 3, 5 and 10 degree.
\subsubsection{Delay Domain}

In the delay domain, a delay scaling factor $S_{dmc}^{\tau}>1$ describing the geometrical extension of DMCs is introduced, readers can find more description from \cite{[DMC3]}.
The delays of LOS component, SMCs and DMCs are calculated as
\begin{equation}
{\tau ^{LOS}}(t) = {\left\| {{\bf{D}}_{pq}^{LOS}(t)} \right\|}/c
\end{equation}
\begin{equation}\label{tau_mn}
{\tau_{s_n}^{SMC}}(t) = [{\left\| {{\bf{D}}_{s_n}^R(t)} \right\|}+{\left\| {{\bf{D}}_{s_n}^T(t)} \right\|}]/c+\tilde{\tau}_n+{\tau_{s_n}}
\end{equation}
\begin{equation}\label{tau_dmc}
{\tau_{m_n}^{DMC}}(t) = {\tau_{s_n}^{SMC}}(t)+{\tau_{add}}
\end{equation}
where $\left\|\cdot\right\|$ calculates the Frobenius norm, $c$ is the speed of light, the virtual link delay $\tilde{\tau}_n$ is calculated as  $\tilde{\tau}_n=-r_{\tau}\sigma_{\tau}ln{\mu_n}$. The random variable $\mu_n\thicksim U(0,1)$, $r_{\tau}$ is the delay scalar and can be found in \cite{[26]} for industrial scene. $\sigma_{\tau}$ is RMS delay spread which obeys log-normal distribution. The mean and variance of $\sigma_{\tau}$ are listed in Table \ref{tab2}.

In (\ref{tau_mn}), the variable $\tau_{s_n}$ is the relative delay of $S_{s_n}$ due to distance offset in \cite{[29]} and obeys exponential distribution with $E[\tau_{s_n}]$ mean. However, it was noted that the delays of the DMCs are not only attributed to the scatterers that are directly visible to the BS and MS, but also to the scattering phenomenon caused by those scatterers that are hidden between the visible scatterers \cite{[DMC3]}. Thus in (\ref{tau_dmc}), an additional delay $\tau_{add}$ from the scatterers that lie between the visible clusters is added at the basis of (\ref{tau_mn}). It is computed as $\tau_{add}=\xi S_{dmc}^{\tau} \beta_{dmc}$, where $\xi\thicksim U(0,1)$, where $\beta_{dmc}$ is the power decay factor.
\subsubsection{Power Domain}

The power of SMC $S_{s_n}$ in cluster $C_n$ is calculated as
\begin{equation}
     \begin{split}
\tilde{P}_{s_n}^{SMC}(t)=\textrm{exp}(-\tau_{s_n}^{SMC}(t)\frac{r_\tau-1}{{r_\tau}{\sigma_\tau}})10^{-Z_n/10}
     \end{split}
\end{equation}
where $r_{\tau}$ is the delay scalar, $\sigma_{\tau}$ is the delay spread, $Z_n\thicksim N(0,\sigma_{cluster})$ and $\sigma_{cluster}$(dB) is the per cluster shadowing standard deviation. The power is normalized as $P_{s_n}^{SMC}(t)={\tilde{P}_{s_n}^{SMC}(t)}/{\sum_{n=1}^{N(t)}\sum_{s=1}^{S_n(t)}\tilde{P}_{s_n}^{SMC}(t)}$.
Finally, the power of DMC $S_{m_n}$ in cluster $C_n$ is calculated as
\begin{equation}
     \begin{split}
\tilde{P}_{m_n}^{DMC}(t)=&max\{P_{s_n}^{SMC}(t)\}\\
&\times P_{off}\textrm{exp}(-\frac{{\tau_{m_n}^{DMC}}(t)-{\tau_{s_n}^{SMC}}(t)}{\beta_{dmc}})
     \end{split}
\end{equation}
where $max\{P_{s_n}^{SMC}(t)\}$ is the power of the strongest SMC in $C_n$, $P_{off}$ is the power offset, the power delay factor $\beta_{dmc}$ represents the excess delay relative to the strongest SMC that attenuates the DMC power to $1/e$ relative to their base power $P_{s_n}(t)P_{off}$.  

\subsubsection{DMC Power Ratio and K Factor}
One of the important indicators in IIoT is the DMC power ratio, which represents the relative power attributable to the DMC part of the radio channel in the industrial scenario. This DMC power ratio can be written as
\begin{equation}
\eta_{dmc}(t)=\frac{P_{DMC}(t)}{P_{DMC}(t)+P_{SMC}(t)+P_{LOS}(t)}
\end{equation}
where $P_{DMC}(t)$ and $P_{SMC}(t)$ are the total power of DMCs and SMCs, respectively. It is known that the power ratio of LOS and NLOS equals to $K$ factor. Assume that the DMC power ratio and $K$ factor are constants for simplification, the power of LOS satisfies $P_{LOS}(t)=K(P_{SMC}(t)+P_{DMC}(t))$. Since the power of SMC have been normalized and the DMC power ratio is assumed to be time-invariant for simplify, we can easily get that the total power of DMCs is $P_{DMC}=\frac{(1+K)\eta_{dmc}}{1-(K+1)\eta_{dmc}}$. Thus the DMC power will be normalized by $P_{m_n}^{DMC}(t)=P_{DMC}P_{n}^{SMC}(t)\cdot\tilde{P}_{m_n}^{DMC}(t)/\sum_{m_n=1}^{M_n(t)}\tilde{P}_{m_n}^{DMC}(t)$, where $P_{n}^{SMC}(t)=\sum_{s=1}^{S_n(t)}\tilde{P}_{s_n}^{SMC}(t)$.
Note that the DMC power ratio is also related to the polarization of the transmitter and receiver\cite{[IF1]}, which can be further studied in the future.

Besides, in order to support non-stationary channels, the time and array evolution of clusters are also considered. A time-array evolution of clusters is introduced in \cite{[29]}. In this paper, the visibility of clusters to different antennas when initializing is considered. Taking Rx as an example, the visible probability of a cluster to receive antennas is $P_T(\Delta Rx)=\exp(-\lambda_R {\Delta Rx\cdot \lambda_c}/{D_c^s})$,
where $\lambda_R$ is the generation rate of clusters, $\Delta Rx$(m) is the unit antenna spacing of Rx array, $\lambda_c$ is the wavelength and $D_c^s$ is a scenario-dependent coefficient describing space correlation.

\subsection{Model Parameters}
3GPP includes values of some model parameters in the indoor factory such as the delay scalar $r_{\tau}$, the mean and variance of $\sigma_{\tau}$, the number of clusters and so on. For the parameters of DMC model, the power offset $P_{off}$ and the DMC power ratio $\eta_{dmc}$ are assumed to be constants for simplification. We choose the typical value of the power offset\cite{[DMC3]} and the DMC power ratios in LOS and NLOS from measurement\cite{[IF2]}. 
The power decay factor and delay scaling factor are set different values respectively to study their impacts on the channel response. Those model parameters are listed in Table \ref{tab2}. Other parameters are determined by the measurement environment and setup, such as the mean of initial distances between antennas and clusters, speeds of antennas and scatterers. The standard variances are obtained by parameter estimation based on the minimum mean square error (MMSE) criterion, which can be found in \cite{[29]}.
\begin{table}[tb]
	\caption{Parameters from Standard Channel Model.}
	\centering
		\begin{tabular}{|m{2.5cm}<{\centering}|m{2.1cm}<{\centering}|m{2.1cm}<{\centering}|}
			\hline
			\textbf{Parameters} & \textbf{LOS}&\textbf{NLOS}\\
             \hline
			$r_\tau$& 2.7& 3\\
            \hline
			$\sigma_{cluster}$ (dB) & 4& 3\\
            \hline
            $\tilde{\lambda}_{smc}$ &3&3\\
            \hline
            $\tilde{\lambda}_{dmc}$ &17&17\\
            \hline
            $E[\sigma_\tau](log_{10}[s])$&-7.53 &-7.41\\
               \hline
               $std[\sigma_\tau](log_{10}[s])$&0.12&0.13\\
             \hline
             $P_{off}$ (dB)&10&10\\
             \hline
             $S_{dmc}^{\tau}$&2(10)&2(10)\\
             \hline
             $\beta_{dmc}$ (ns)&10(50)&10(50)\\
             \hline
             $\eta_{dmc}$&0.14&0.4\\
             \hline
		\end{tabular}
		\label{tab2}
\end{table}
\section{Statistical Property Analysis}
\subsection{Time-varying Transfer Function}
The time-varying transfer function is calculated as the Fourier transform of channel impulse response (\ref{tf}), where $f$ is the center frequency.
\begin{equation}\label{tf}
\begin{aligned}
H_{qp}(t&,f)=\int_{-\infty}^{\infty}h_{qp}(t,\tau)e^{-j2\pi f\tau}d\tau\\
=&\sqrt{\frac{K}{K+1}}h_{qp}^{LOS}(t)e^{-j2\pi f\tau^{LOS}(t)}\\
+&\sqrt{\frac{1}{K+1}}\sum_{n=1}^{N(t)}\sum_{s_n=1}^{S_n(t)}h_{qp,{s_n}}^{SMC}(t)e^{-j2\pi f\tau_{s_n}^{SMC}(t)}\\
+&\sqrt{\frac{1}{K+1}}\sum_{n=1}^{N(t)}\sum_{m_n=1}^{M_n(t)}h_{qp,m_n}^{DMC}(t)e^{-j2\pi f\tau_{m_n}^{DMC}(t)}
\end{aligned}
\end{equation}
\subsection{Space-Time-Frequency Correlation Function}
In order to study the correlation of the industrial channel, the STFCF of proposed model can be derived from the time-varying transfer function. Assuming that the LOS and NLOS components are uncorrelated, it is calculated as
\begin{equation}\label{stfcf}
\begin{split}
{R_{qp,q'p'}}&({\delta _T},{\delta _R},\Delta f,\Delta t;t,f)\\
=& E[{H_{qp}}(t,f)H_{q'p'}^*(t + \Delta t,f + \Delta f)]\\
=&{R_{qp,q'p'}^L}({\delta _T},{\delta _R},\Delta f,\Delta t;t,f)\\
+&{R_{qp,q'p'}^N}({\delta _T},{\delta _R},\Delta f,\Delta t;t,f)
\end{split}
\end{equation}
where the STFCF is time-variant, frequency-variant and related to the antenna spacing ${\delta _T} = \left\| {{\bf{A}}_p^T - {\bf{A}}_{p'}^T} \right\|$, ${\delta _R} = \left\| {{\bf{A}}_q^R - {\bf{A}}_{q'}^R} \right\|$, time difference $\Delta t$ and frequency difference $\Delta f$, but for simplification, the different correlation parts $R_{qp,q'p'}^{X}({\delta _T},{\delta _R},\Delta f,\Delta t;t,f)$ are denoted as $R_{qp,q'p'}^{X}$, where $X=L,N$ represent LOS and NLOS, respectively. They are computed as
\begin{equation}
\begin{split}
{R_{qp,q'p'}^L}
=\frac{K}{K+1}&h_{qp}^{LOS}(t)h_{q'p'}^{LOS*}(t+\Delta t)e^{j2\pi\sigma_{0}}
\end{split}
\end{equation}
\begin{equation}\label{STFCF_N}
\begin{split}
&{R_{qp,q'p'}^N}={R_{qp,q'p'}^{SS'}}+{R_{qp,q'p'}^{SM'}}
+{R_{qp,q'p'}^{MS'}}
+{R_{qp,q'p'}^{MM'}}
\end{split}
\end{equation}
where $\sigma_0=f[\tau^{LOS}(t)-\tau^{LOS}(t+\Delta t)]+\Delta f\tau^{LOS}(t+\Delta t)$, the STFCF of NLOS components consists of four parts, as shown in (\ref{STFCF_4parts}), where $R_{qp,q'p'}^{SM'}$ represents the correlation between the SMC at instant $t$, center frequency $f$, with $q,p$ antenna pair and the DMC at instant $t+\Delta t$, center frequency $f+\Delta f$, with $q',p'$ antenna pair. The other three are similar. 
\begin{figure*}[b]
	{\noindent} \rule[-3pt]{18.07cm}{0.06em}
	\begin{subequations}\label{STFCF_4parts}
		\begin{align}
&{R_{qp,q'p'}^{SS'}}=\frac{1}{K+1}E[\sum_{n=1}^{N(t)}\sum_{s_n=1}^{S_n(t)}\sum_{n'=1}^{N(t+\Delta t)}\sum_{s_{n'}=1}^{S_{n'}(t+\Delta t)}h_{qp,s_n}^{SMC}(t)
h_{q'p',s_{n'}}^{SMC*}(t+\Delta t)e^{j2\pi\sigma_{1}}]\\
&{R_{qp,q'p'}^{SM'}}=\frac{1}{K+1}E[\sum_{n=1}^{N(t)}\sum_{s_n=1}^{S_n(t)}\sum_{n'=1}^{N(t+\Delta t)}\sum_{m_{n'}=1}^{M_{n'}(t+\Delta t)}h_{qp,s_n}^{SMC}(t)h_{q'p',m_{n'}}^{DMC*}(t+\Delta t)e^{j2\pi\sigma_{2}}]\\
&{R_{qp,q'p'}^{MS'}}=\frac{1}{K+1}E[\sum_{n=1}^{N(t)}\sum_{m_n=1}^{M_n(t)}\sum_{n'=1}^{N(t+\Delta t)}\sum_{s_{n'}=1}^{S_{n'}(t+\Delta t)}h_{qp,m_n}^{DMC}(t)
h_{q'p',s_{n'}}^{SMC*}(t+\Delta t)e^{j2\pi\sigma_{3}}]\\
&{R_{qp,q'p'}^{MM'}}=\frac{1}{K+1}E[\sum_{n=1}^{N(t)}\sum_{n'=1}^{N(t+\Delta t)}\sum_{m_n=1}^{M_n(t)}\sum_{m_{n'}=1}^{M_{n'}(t+\Delta t)}h_{qp,m_n}^{DMC}(t)h_{q'p',m_{n'}}^{DMC*}(t+\Delta t)e^{j2\pi\sigma_{4}}]
		\end{align}
	\end{subequations}
\end{figure*}
where $\sigma_1=f[\tau_{s_n}^{SMC}(t)-\tau_{s_{n'}}^{SMC}(t+\Delta t)]-\Delta f\tau_{s_{n'}}^{SMC}(t+\Delta t)$, $\sigma_2=f[\tau_{s_n}^{SMC}(t)-\tau_{m_{n'}}^{DMC}(t+\Delta t)]-\Delta f\tau_{m_{n'}}^{DMC}(t+\Delta t)$, $\sigma_3=f[\tau_{m_n}^{DMC}(t)-\tau_{s_{n'}}^{SMC}(t+\Delta t)]-\Delta f\tau_{s_{n'}}^{SMC}(t+\Delta t)$,  $\sigma_4=f[\tau_{m_n}^{DMC}(t)-\tau_{m_{n'}}^{DMC}(t+\Delta t)]-\Delta f\tau_{m_{n'}}^{DMC}(t+\Delta t)$.

Finally, STFCF can be simplified to temporal autocorrelation function (ACF), spacial cross-correlation function (CCF) and frequency correlation function (FCF) \cite{[29]}.

\subsection{Time-variant RMS Delay Spread}
The time-variant RMS delay spread is calculated as
\begin{equation}\label{ds}
\begin{split}
\sigma_\tau(t)=\sqrt{\frac{\sum P(\tau_l(t))\tau_l^2(t)}{\sum P(\tau_l(t))}-(\frac{\sum P(\tau_l(t))\tau_l(t)}{\sum P(\tau_l(t))})^2}
\end{split}
\end{equation}
where $\tau_l(t)$ is the delay of the $l$-th path at $t$ instant, $P(\tau_l(t))$ is the power of the corresponding path.
\section{Results and Analysis}
\begin{figure}[tb]
	\centerline{\includegraphics[width=0.38\textwidth]{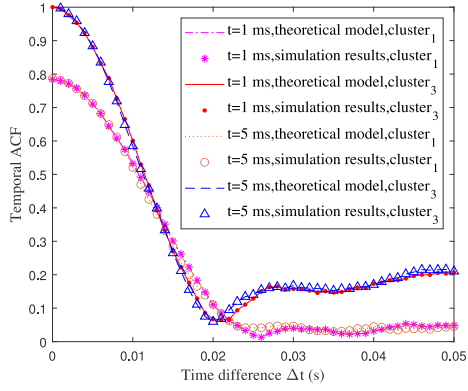}}
	\caption{Comparison of the normalized analytical and simulated ACFs of $Cluster_1$ and $Cluster_3$ ($f_c$=5.8 GHz, $D$=1 m, $M_R$=1, $M_T$=1, $|v^R|$=1 m/s, $|v^T|$=0 m/s, $D_c^s$=100 m, $P_{off}$=10 dB, $\beta_{dmc}$=10 ns, $S_{dmc}^{\tau}$=2, NLOS).}
	\label{fig_3}
\end{figure}

This section introduces the simulation results of the model. We choose an indoor factory scenario at 5.8 GHz radio frequency in \cite{[29]} to verify it. 
\subsubsection{ACF}
Fig.~\ref{fig_3} presents the simulation and theoretical temporal ACFs of $\textrm{cluster}_1$ and $\textrm{cluster}_3$ at different instants in NLOS condition. Tx and Rx are linear arrays with half wavelength spacing. All the ACFs are normalized relative to $\textrm{cluster}_3$, and the simulation parameters are listed under the figure.
In our model, the close solution of STFCF cannot be derived due to a large amount of random variables, so the theoretical results are the realization of raw equations by transfer function. Instead, the simulation results are calculated using channel coefficients assuming that LOS and NLOS components are uncorrelated.
It can be seen that all the simulation temporal ACFs have a good consistency with the theoretical results, validating the correctness of the model. Besides, we can observe different ACFs at $t=$1 $ms$ and $t=5$ $ms$, which are resulted from time-varying angles, distances, Tx, Rx and clusters’ locations in the proposed model, illustrating that our model can mimic the non-stationarity of industrial channels.
\begin{figure}[tb]
	\centerline{\includegraphics[width=0.53\textwidth]{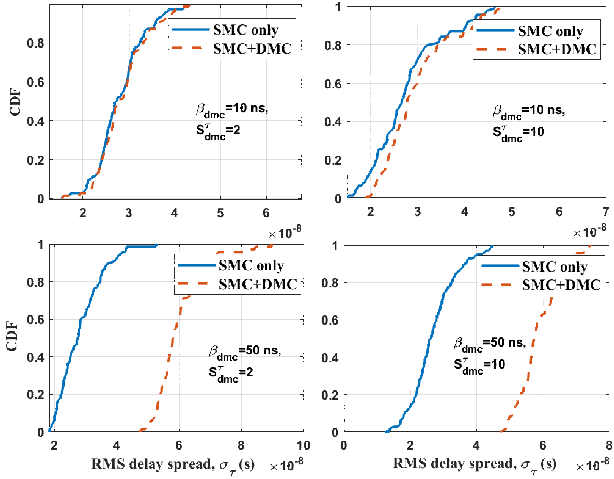}}
	\caption{Comparison of RMS delay spread of SMC only and SMC+DMC with different delay scaling factors and power decay factors ($f_c$=5.8 GHz, $D$=1~m, $M_R$=1, $M_T$=1, $|v^R|$=0.4 m/s, $|v^T|$=0 m/s, $D_c^s$=100 m, $P_{off}$=10~$dB$, $\eta_{dmc}=0.4$, NLOS).}
	\label{fig_4}
\end{figure}
\subsubsection{DMC Power Modeling Analysis}
The properties of DMC and relevant parameters are also studied. Fig.~\ref{fig_4} shows the CDF of RMS delay spread of SMC only and SMC+DMC with different delay scaling factors and power decay factors. Each individual picture shows that DMCs increase the RMS delay spread on average, which is consistent with the conclusion in \cite{[DMC3]}. Different from \cite{[DMC3]}, it can be seen that $\beta_{dmc}$ increases the delay spread obviously from the top two and the bottom two pictures in Fig.~\ref{fig_4}. This is because DMCs account for a large part of the total power in IIoT, the impact of DMCs on the delay spread is non-negligible. A larger $\beta_{dmc}$ produces larger power at larger delay, leading to larger delay spread. However, comparing the left two and the right two pictures, it can be observed that the delay scaling factor does not have much effect on the RMS delay spread. This is because although the access delay of DMC increases with the $S_{dmc}^{\tau}$, the relative power of DMC also decreases. What is more, the simulation results show that $\beta_{dmc}$ and $S_{dmc}^{\tau}$ do not have a significant effect on the delay spread in LOS condition, for the reason that the LOS component is strong enough to determine it.


\begin{figure}[tb]
	\centerline{\includegraphics[width=0.43\textwidth]{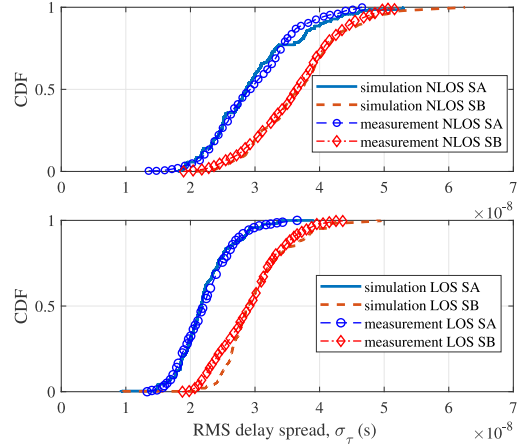}}
	\caption{Comparison of the simulation and measurements of CDFs of delay spread in Scenario A and B NLOS ($f_c$=5.8 GHz, $D$=1 m, $M_R$=1, $M_T$=1, [std($\phi_n^A$), std($\phi_n^E$), std($\varphi_n^A$), std($\varphi_n^E$)]=[31.8, 16, 30.6, 10.2], $|v^R|$=0.4 m/s, $|v^T|$=0 m/s, $S_{dmc}^{\tau}$=2, $\beta_{dmc}$=10 ns, $K$=11 dB in SA,$K$ =7 dB in SB).}
	\label{fig_6}
\end{figure}

\subsubsection{Comparison with Measurements}
3GPP gave four IIoT sub-scenarios according to the antenna height and clutter density\cite{[3GPP],[26]}. Because the antenna height mainly has an impact on path loss and LOS probability, and there has not been relevant measurement data in four IIoT sub-scenarios, we choose the measurement data in two sub-scenarios, light clutter (Scenario A, SA) and heavy clutter (Scenario B, SB) environment\cite{[6]}. Considering LOS and NLOS, four circumstances are measured respectively.
According to 3GPP, the number of clusters in SB is lager than that in SA.

For the specific industrial environment, the distance between Tx and Rx, the moving speed of Tx, Rx and scatterers, antenna patterns and so on are set to be consistent with the measurement. Parameters like angular standard deviations and K factor are estimated based on MMSE criterion. Fig.~\ref{fig_6} represents the comparison of CDFs of delay spread between the simulation and measurements in four conditions. The simulation parameters are listed below the figures. It can be seen that the simulation can be well matched with the measurement results. In general, the delay spread in NLOS scenario is larger than that in LOS scenario, and delay spread in SB is larger than that in SA. Note that the K factor in the clutter-heavy environment is smaller, the reason is that more multipath components result in worse channel.

\section{Conclusions}
In this paper, a 3D non-stationary GBSM which supports massive MIMO, millimeter wave communication and the rich DMCs in IIoT has been proposed for industrial automation wireless communication systems and proved to be applicable to different industrial sub-scenarios. The statistical properties of the proposed channel model have been investigated to demonstrate its capability of capturing channel characteristics of industrial environments, with excellent fitting to some corresponding channel measurements. In the future, industrial channel measurements in four sub-scenarios in 3GPP are expected to be completed. Besides, the polarization related DMC power ratio is also expected to be considered.
\section*{Acknowledgment}
\small {This work was supported by the National Key R\&D Program of China under Grant 2018YFB1801101, the National Natural Science Foundation of China (NSFC) under Grant 61960206006, the Frontiers Science Center for Mobile Information Communication and Security, the High Level Innovation and Entrepreneurial Research Team Program in Jiangsu, the High Level Innovation and Entrepreneurial Talent Introduction Program in Jiangsu, the Research Fund of National Mobile Communications Research Laboratory, Southeast University, under Grant 2020B01, the Fundamental Research Funds for the Central Universities under Grant 2242021R30001, and the EU H2020 RISE TESTBED2 project under Grant 872172.}


\end{document}